\documentclass{aastex}

\usepackage{epsf}
\usepackage{emulateapj5}

\usepackage{graphicx}
\usepackage{amsmath}
\usepackage{times}
\usepackage{colordvi}

\shorttitle{A powerful hydrodynamic booster for relativistic jets}
\shortauthors{Aloy \& Rezzolla}

\begin{document}

\title{A powerful hydrodynamic booster for relativistic jets}

\author{Miguel A. Aloy\altaffilmark{1,~}\altaffilmark{{2}} and Luciano
Rezzolla\altaffilmark{3,~}\altaffilmark{4,~}\altaffilmark{5}}

\altaffiltext{1}{Max--Planck--Institut f\"ur Astrophysik,
  {Karl-Schwarzschild-Str.~1, 85741} Garching, Germany}

\altaffiltext{2}{Departamento de Astronom\'{\i}a y Astrof\'{\i}sica,
  Universidad de Valencia, {C/. Doctor Moliner, s/n, 46100}
  Burjassot, Spain}

\altaffiltext{3}{Max--Planck--Institut f\"ur Gravitationsphysik,
Albert Einstein Institut, {Am Muehlenberg 1, 14476} Golm, Germany}

\altaffiltext{4}{SISSA, International School for Advanced Studies and
INFN, Via Beirut, 2 34014 Trieste, Italy}

\altaffiltext{5}{Department of Physics and Astronomy, Louisiana State
University, 202 Nicholson Hall, Baton Rouge, LA 70803, USA}

\date{\today}

\begin{abstract}
  Velocities  close to  the speed  of  light are  a robust  observational
  property  of  the jets  observed  in  microquasars  and AGNs,  and  are
  expected  to be behind  much of  the phenomenology  of GRBs.   Yet, the
  mechanism boosting  relativistic jets to such large  Lorentz factors is
  still  essentially unknown.   Building on  recent general-relativistic,
  multidimensional  simulations of  progenitors of  short  {GRBs}, we
  discuss a new effect in  relativistic hydrodynamics which can act as an
  efficient booster  in jets.  This  effect is purely  hydrodynamical and
  occurs when large velocities  tangential to a discontinuity are present
  in the flow, yielding Lorentz factors $\Gamma \sim 10^2-10^3$ or larger
  in  flows with moderate  initial Lorentz  factors.  Although  without a
  Newtonian counterpart, this effect  can be explained easily through the
  most elementary hydrodynamical flow:  {\em i.e.} a relativistic Riemann
  problem.
\end{abstract}


\maketitle

\section{INTRODUCTION}
\label{intro}

        Determining the mechanism accelerating high-energy plasma to the
relativistic velocities observed in jets from Active Galactic Nuclei
(AGNs) and microquasars, and expected in gamma-ray bursts (GRBs) remains
a problem whose solution has long been sought in any consistent modelling
of these astrophysical objects. We here show how recent
general-relativistic simulations and novel effects of relativistic
hydrodynamics can provide important clues in this search.

In crude but not oversimplified terms, the dynamics of a relativistic
hydrodynamic jet in an external medium can be assimilated to the motion
of two fluids, one of which is much hotter and at higher (or equal)
pressure than the other one, and is moving with a large tangential
velocity with respect to the cold, slowly moving fluid. In this scenario,
a blob of hot plasma moving outwards along the jet generically produces a
hydrodynamical structure, in the direction perpendicular to the motion,
composed of a ``forward'' shock (moving away from the jet-axis) and of a
``reverse'' shock (moving towards the jet-axis). This pattern of waves,
which we indicate as ${\cal _{\leftarrow}\!S C S_{\!\rightarrow}}$, where
${\cal _{\leftarrow}\!S}$ refers to the reverse shock, ${\cal
S_{\!\rightarrow}}$ to the forward one and ${\cal C}$ to the contact
discontinuity between the two, is routinely produced in numerical
simulations of relativistic jets. Surprisingly however, because
unexpected from Newtonian hydrodynamics, this pattern can change for
sufficiently large tangential velocities and in this case the
inward-moving shock is replaced by a rarefaction wave thus producing a
${\cal _{\leftarrow}\!R C S_{\!\rightarrow}}$ pattern.  Convincing
evidence that this process is likely to occur under realistic conditions
is now emerging from special relativistic calculations of extragalactic
jets~\citep{Alo03}, as well as from general relativistic,
multidimensional simulations of progenitors of short
GRBs~\citep{Alo05}. The latter case is schematically shown in
Fig.~\ref{fig:GRBscheme} {which shows the flow-structure produced in
numerical simulations involving a thick accretion torus around a black
hole and in which a ultrarelativistic outflow is generated.}

        Because disentangling this change in the wave-pattern, first
pointed out by~\cite{RZ02}, and understanding its influence on the
complex dynamics of a relativistic jet can be very difficult, we resort
to the simplest hydrodynamical flow in which this effect can occur: {\em
i.e.,} a Riemann problem.

\section{The relativistic boosting mechanism}
\label{trbm}

        We recall that a Riemann problem consists of determining the
evolution of a fluid which is composed at some initial time by two
uniform states ({\em i.e.,} a {``left'' and a ``right''} state) with
different and discontinuous hydrodynamical properties: the rest-mass
density $\rho$, the pressure $p$, the specific internal energy
$\varepsilon$, the specific enthalpy $h \equiv 1+\varepsilon/c^2+p/(\rho
c^2)$ and the components of the velocity normal $v^n$ and tangential
$v^t$ to the initial discontinuity. Using this initial setup and the
equations of relativistic hydrodynamics,~\cite{RZP03} revealed that, in
contrast with what happens in Newtonian hydrodynamics, a smooth
transition from one wave-pattern to another can be produced by simply
varying the velocities which are tangential to the initial discontinuity
$v^t$, while maintaining the initial states unmodified. More
specifically, for each set of initial states there exists a critical
value for the initial tangential velocities at which a shock is
transformed into a rarefaction wave or viceversa.

The discovery of this relativistic effect is not purely academic and
without direct physical implications. Indeed, an important consequence
of this change of waves is the possibility of producing a large
amplification of the local the Lorentz factor $\Gamma \equiv [1 -
(v^t)^2 - (v^n)^2]^{-1/2}$ when, past a critical value for the
tangential velocity, a shock wave transforms into a rarefaction wave.
To illustrate how this can happen in practice it useful to consider a
Riemann problem with initial states that would mimic the conditions
for the generation of a GRB or the propagation of a plasma-blob along
a relativistic jet (either extragalactic or in a microquasar). This
can be done in practice by considering a left state which is much
hotter, with larger pressure and smaller rest-mass density than the
right state. Stated differently, this corresponds to a left state
having a much larger specific enthalpy than the right one. Such a
choice is justified by the fact that the jet is expected to be very
underdense with respect to the external medium (which {could be given,
  for instance, by} a dense accretion torus) and to be overpressured
because a substantial change in the jet radius must happen between
{its} {launching} site and the location at which it is detected $z$.
For GRBs, for instance, one expects $z \approx 10^{13}-10^{14}\,$cm,
while $z \approx 10^{16}-10^{17}\,$cm for microquasars and $z \approx
10^{18}\,$cm for extragalactic jets.

\vbox{
\vskip 0.125truecm
\centerline{\epsfxsize=8.truecm\epsfbox{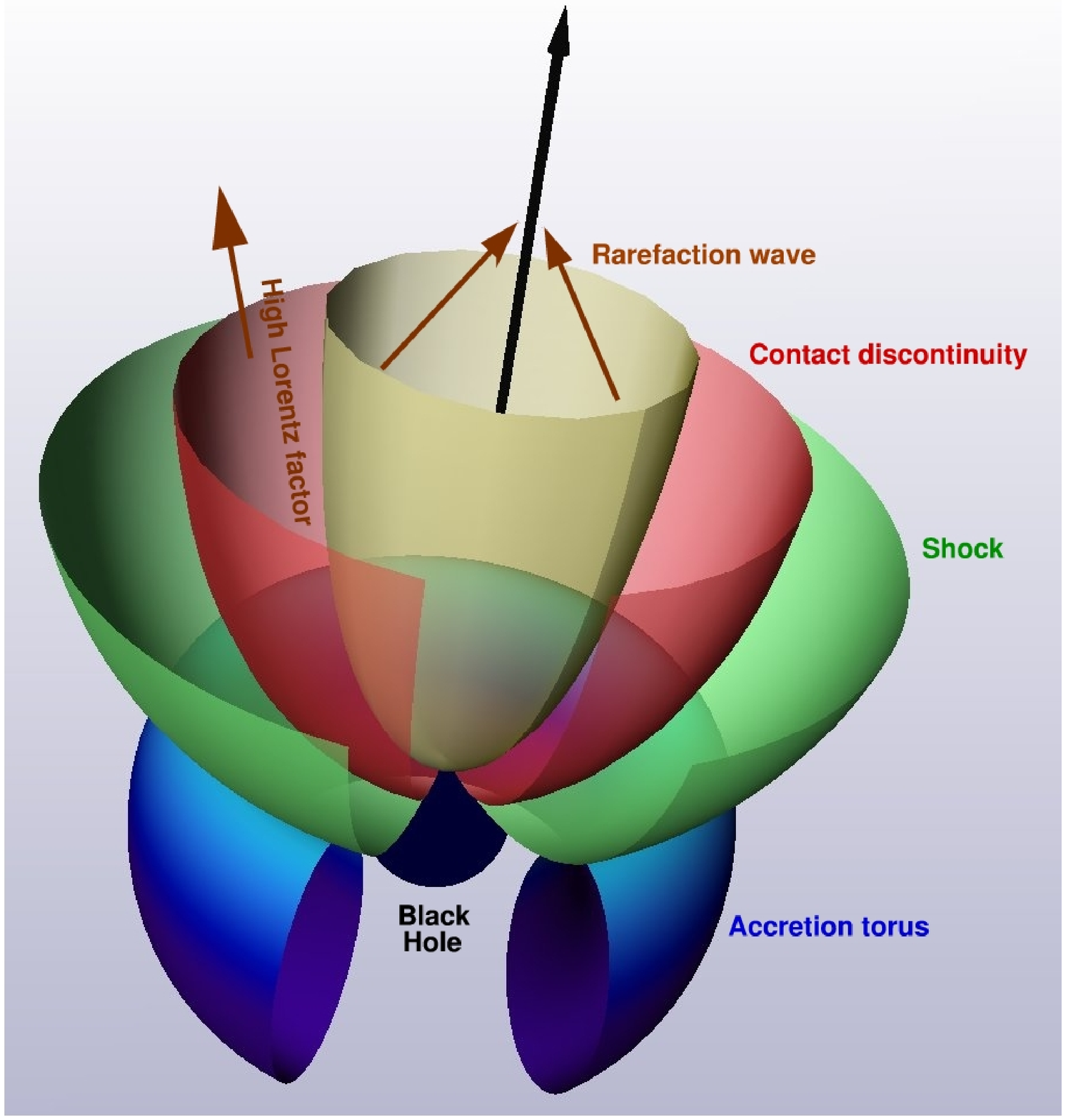}}

\figcaption[]{\label{fig:GRBscheme}Schematic illustration of the flow
  structure in the acceleration of a short GRB produced by an accretion
  torus orbiting around a {stellar-mass} black hole.  The arrows mark the
  direction of fluid velocity at the rarefaction head (yellow
  paraboloidal surface), indicating that collimation of the fluid tends
  towards the black-hole rotation axis (black arrow). A large boost is
  produced in the region between the rarefaction head and the contact
  discontinuity (red surface) separating the relativistic outflow from
  the shocked external medium.}  \vskip 0.125truecm }

Keeping these initial states fixed, we can then use the tangential
velocity in the left state $v^t_{_{\rm L}}$ as a free parameter, whose
tuning allows to go from one wave-pattern to the other. More
specifically, let us consider the right state to be a cold fluid with
a large rest-mass density and essentially at rest, {\em i.e.,} with
initial hydrodynamical properties: \hbox{$p_{_{\rm R}}\approx
  10^{-6}\rho_{_{\rm ext}} c^2$}, \hbox{$\rho_{_{\rm R}} \approx
  10^{-2}\rho_{_{\rm ext}}$}, \hbox{$v^n_{_{\rm R}} \approx 0$},
\hbox{$v^t_{_{\rm R}} \approx 0$}, where $\rho_{_{\rm ext}}$ is a
normalization constant which allows us to be scale-free.  For
convenience, herafter we will assume $c = 1$ and provide explicit
values for $\rho_{_{\rm ext}}$ when discussing the astrophysical
implications in Sect. 3. The left state will have comparable but
larger pressure, will be underdense, with a small initial tangential
velocity and a normal one close to the speed of light, {\em i.e.,}
\hbox{$p_{_{\rm L}} \gtrsim \rho_{_{\rm R}}$}, \hbox{$\rho_{_{\rm L}}
  \ll \rho_{_{\rm R}}$}, \hbox{$v^n_{_{\rm L}} \approx {1}$},
\hbox{$v^t_{_{\rm L}} \approx 0$}.

\begin{figure*}
\hskip 0.9truecm
\includegraphics*[width=8.cm]{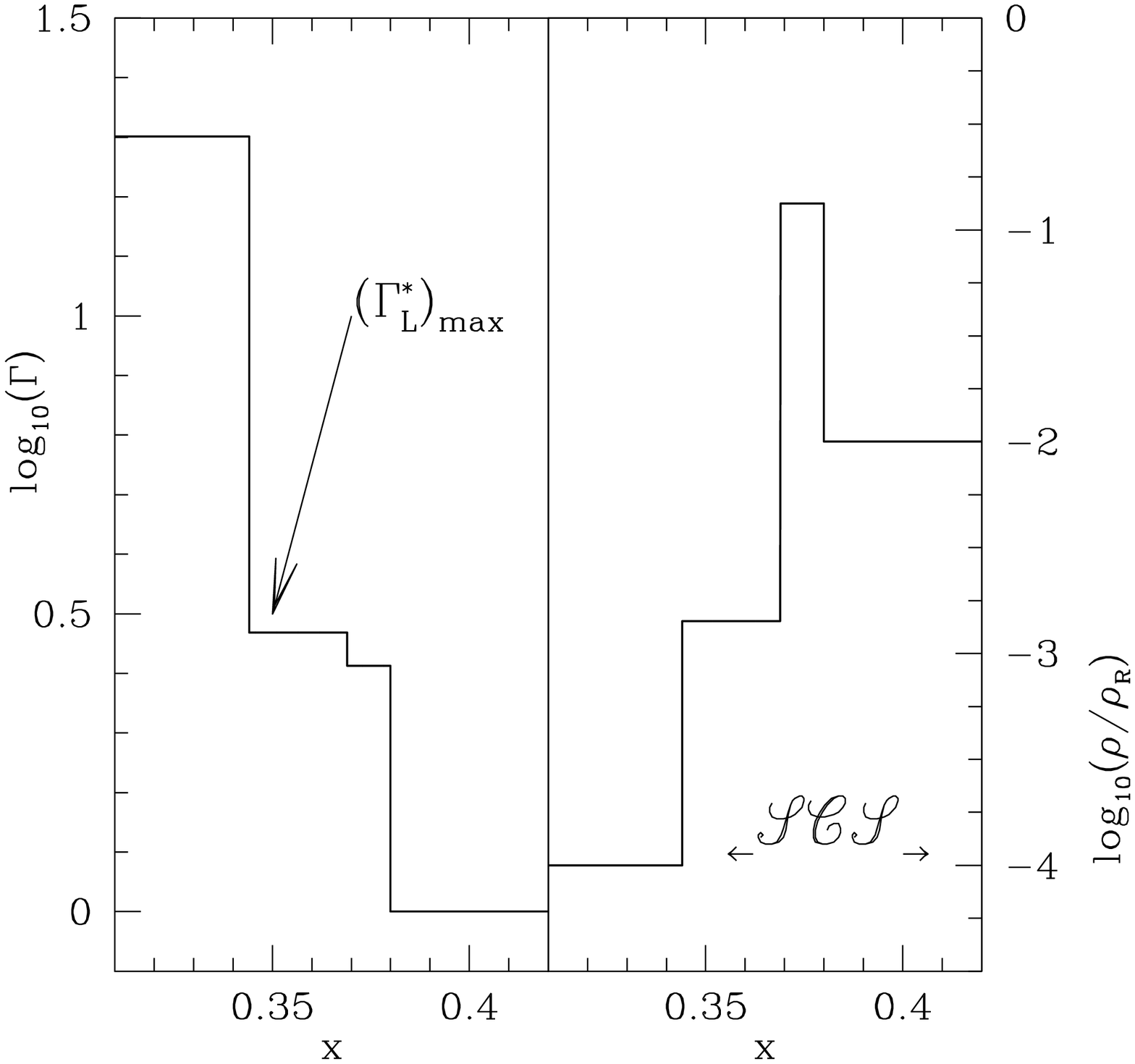}
\hskip 1.truecm
\includegraphics*[width=8.cm]{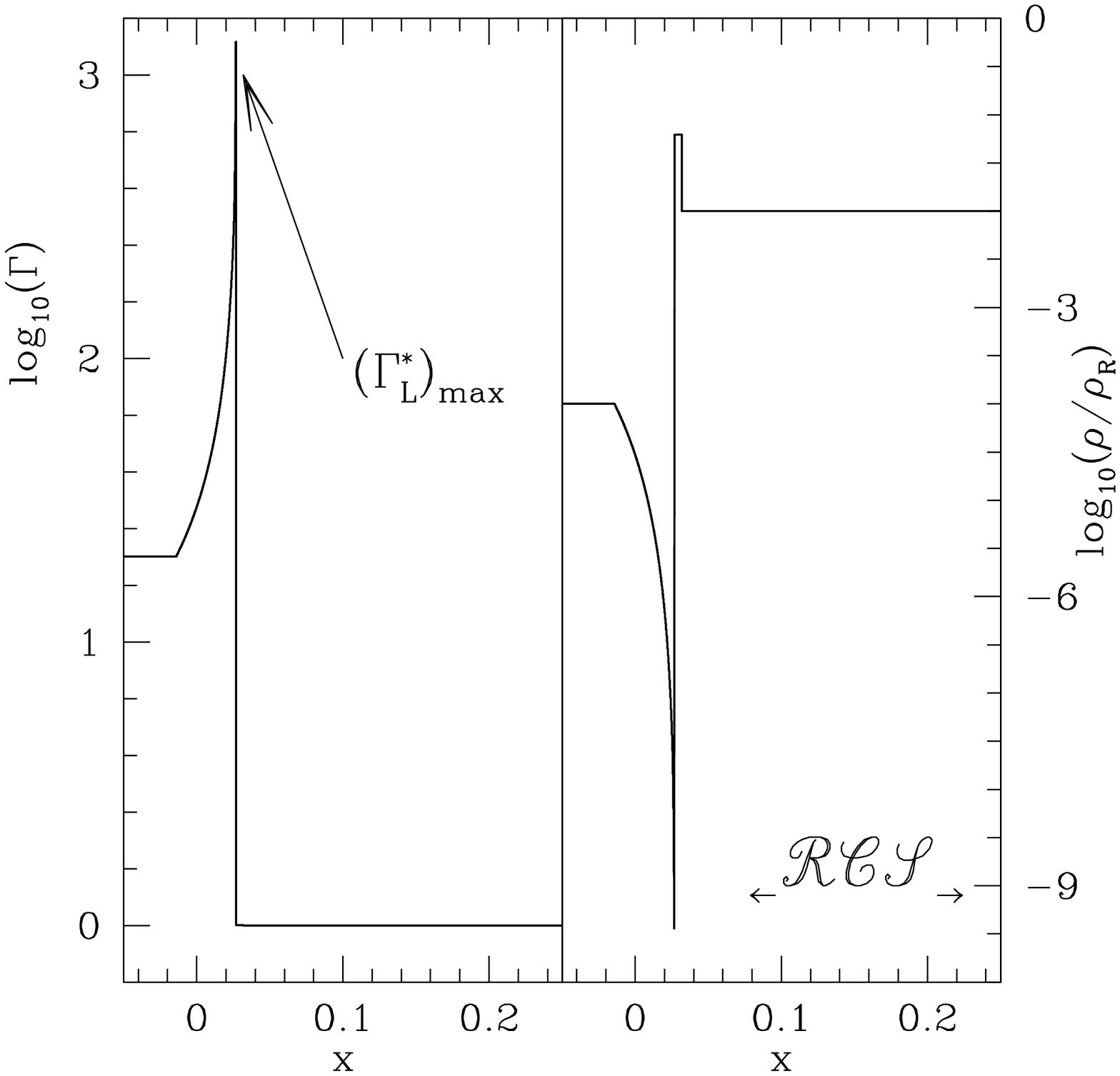}
\caption{\label{fig:scs_rcs} \textit{Left panel:} Profiles for the
  Lorentz factor (left) and rest-mass density (right) resulting from a
  prototypical Riemann problem yielding a ${\cal _{\leftarrow}\!S C
    S_{\!\rightarrow}}$ pattern. The initial states are $(p_{_{\rm
      L}}, \rho_{_{\rm L}}, v^n_{_{\rm L}}, \Gamma_{_{\rm L}}) =
  (10^{-3}, 10^{-4},0.99,20)$ and $(p_{_{\rm R}}, \rho_{_{\rm R}},
  v^n_{_{\rm R}}, \Gamma_{_{\rm R}}) = (10^{-6},10^{-2},0,1)$. Note
  that no {boost} is produced at the contact discontinuity ({\it
    i.e.,} $(\Gamma^*_{_{\rm L}})_{\rm max} < \Gamma_{_{\rm L}}$).
  \textit{Right panel:} Same as in the left panel but for a prototype
  ${\cal _{\leftarrow}\!R C S_{\!\rightarrow}}$ wave-pattern. The
  initial states are $(p_{_{\rm L}}, \rho_{_{\rm L}}, v^n_{_{\rm L}},
  \Gamma_{_{\rm L}}) = (10^3,10^{-4},0,20)$ and $(p_{_{\rm R}},
  \rho_{_{\rm R}}, v^n_{_{\rm R}}, \Gamma_{_{\rm R}}) =
  (10^{-6},10^{-2},0,1)$.  Note in the left panel the very large boost
  ({\it i.e.,} $(\Gamma^*_{_{\rm L}})_{\rm max} \gg \Gamma_{_{\rm
      L}}$) which peaks at the contact discontinuity at $x\approx
  0.027$.}
\end{figure*}

        The solution of the Riemann problem produced by these two states
after an arbitrary time is shown in the left panel of
Fig.~\ref{fig:scs_rcs} (We recall that the solution of the Riemann
problem is self-similar, {\em i.e.,} the structure of the solution in the
different waves does not depend on time.). The solution is of type ${\cal
_{\leftarrow}\!S C S_{\!\rightarrow}}$ and thus consisting of a shock
propagating to the left, of a shock propagating towards the right and of
a contact discontinuity at $x\approx 0.369$ separating the two
waves. This double-shock structure has a well-known Newtonian counterpart
and is commonly observed in numerical simulations when there is an almost
pressure-matching between the jet and the environment. Note that the
acceleration at the transition layer between the jet and the external
medium is very small and that the Lorentz factor reached in the state
between the left-propagating shock and the contact discontinuity,
$\Gamma^*_{_{\rm L}}$, is smaller than the initial Lorentz factor in the
left state $\Gamma_{_{\rm L}}$. This is indicated as $(\Gamma^*_{_{\rm
L}})_{\rm max}$ in Fig.~\ref{fig:scs_rcs} and reaches a value
$(\Gamma^*_{_{\rm L}})_{\rm max} \approx 2.943$ against an initial one in
the left state $\Gamma_{_{\rm L}}=20$.
      
While keeping the right state to be the same, let us now consider the
jet to be considerably over-pressured and with almost no velocity
component perpendicular to the jet direction. The left state can then
be {parameterized} as: \hbox{$p_{_{\rm L}} \gg \rho_{_{\rm R}}$},
\hbox{$\rho_{_{\rm L}} \ll \rho_{_{\rm R}}$}, \hbox{$v^n_{_{\rm L}}
  \approx 0$}, \hbox{$v^t_{_{\rm L}} \approx {1}$}, and the
solution of the Riemann problem in this case is shown in the right
panel of Fig.~\ref{fig:scs_rcs}. This is now of type ${\cal
  _{\leftarrow}\!R C S_{\!\rightarrow}}$ and consists of a rarefaction
wave propagating to the left, of a contact discontinuity at $x \approx
0.027$ and of a shock propagating towards the right. This change in
wave-pattern is purely relativistic and it is accompanied by a large
increase in the Lorentz factor across the rarefaction wave, with the
maximum value $(\Gamma^*_{_{\rm L}})_{\rm max} \approx 10^3$ reached
at the contact discontinuity.

        This is a remarkable feature of relativistic hydrodynamics:
{namely,} the production of a very strong rarefaction near the
contact discontinuity {\it and} the combined conservation across such a
wave of the specific enthalpy and of the Lorentz factor ({\em i.e.,} of
the quantity $h \Gamma$), has the consequence of accelerating the fluid
to ultrarelativistic regimes with $\Gamma \sim 10^4$ or larger. In
essence, this boosting mechanism efficiently converts into kinetic energy
the work done on the jet by the external medium.

{It is worth stressing again that the occurrence of these large fluid
accelerations is not the result of peculiar initial conditions in a
simplified hydrodynamical flow. Rather, accelerations produced through
this effect are a robust feature in numerical simulations of
{relativistic} jets {\citep{Alo03,Alo05}}. In particular, the recent
simulations of \cite{Alo05} show that, in spite of the complex
{multidimensional} dynamics produced by the generation and propagation of
the jet, the basic feature of this relativistic effect is preserved: {\it
i.e.,}} the growth of the Lorentz factor close to the contact
discontinuity separating a high-enthalpy and high-velocity state from a
low-enthalpy one.

{As an illustrative example, we report in Fig.~\ref{fig:GRB} results
from} {an axisymmetric} {simulation} in which the ultrarelativistic jet
is overpressured with respect to the external medium and its velocity is
almost parallel to the jet/external medium interface {(Aloy et al.
2005)}. Shown along the polar $\theta$-direction at a distance of $3.2
\times 10^7\,$cm from the black hole are the logarithm of the rest-mass
density (dashed line) and the profile of the Lorentz factor (solid
line). The latter, in particular, reaches a maximum $\Gamma \approx 29$
in the left-going rarefaction wave moving towards the axis. These
simulations have also revealed that although the interaction between the
jet and the external medium is prone to Kelvin-Helmoltz and shear driven
instabilities~\citep{Alo02}, the growth of the Lorentz factor imprinted
during the initial phases of the jet propagation, {\em i.e.,} when the
jet interacts with the lateral borders of the accretion torus, remains
{preserved in the subsequent evolution of the flow.}

Let us now illustrate in detail how the final fluid acceleration
depends on the initial conditions by assuming that the right state is
held fixed. In this way, in fact, the transition from one wave-pattern
to the other and the boost in the flow depend on two distinct but
related factors: the value of the pressure in the left state $p_{_{\rm
    L}}$ and the value of the tangential velocity there, $v^t_{_{\rm
    L}}$.

        In particular, it is possible to show that for very small
left-state pressures ({\em i.e.,} $p_{_{\rm L}} \lesssim 10^{-6}$ for the
initial states used above) the system evolves to produce a Riemann
structure of the type {${\cal _{\leftarrow}\!S C S_{\!\rightarrow}}$}
also when the normal velocity is zero ({\em i.e.,} $v^n_{_{\rm L}}
\approx 0$). On the other hand, for very large left-state pressures ({\em
i.e.,} $p_{_{\rm L}} > 1$ for initial states used above), the system
evolves to produce a Riemann structure of the type ${\cal
_{\leftarrow}\!R C S_{\!\rightarrow}}$ also when the normal velocity is
ultrarelativistic ({\em i.e.,} $v^n_{_{\rm L}} \approx 1$).  For all the
values of $p_{_{\rm L}}$ within this range, a critical normal velocity
$(v^n_{_{\rm L}})_c$ exists below which the left-propagating shock is
replaced by a rarefaction wave and a large boost takes place.
\vbox{
\vskip 0.125truecm
\centerline{\epsfxsize=8.truecm\epsfbox{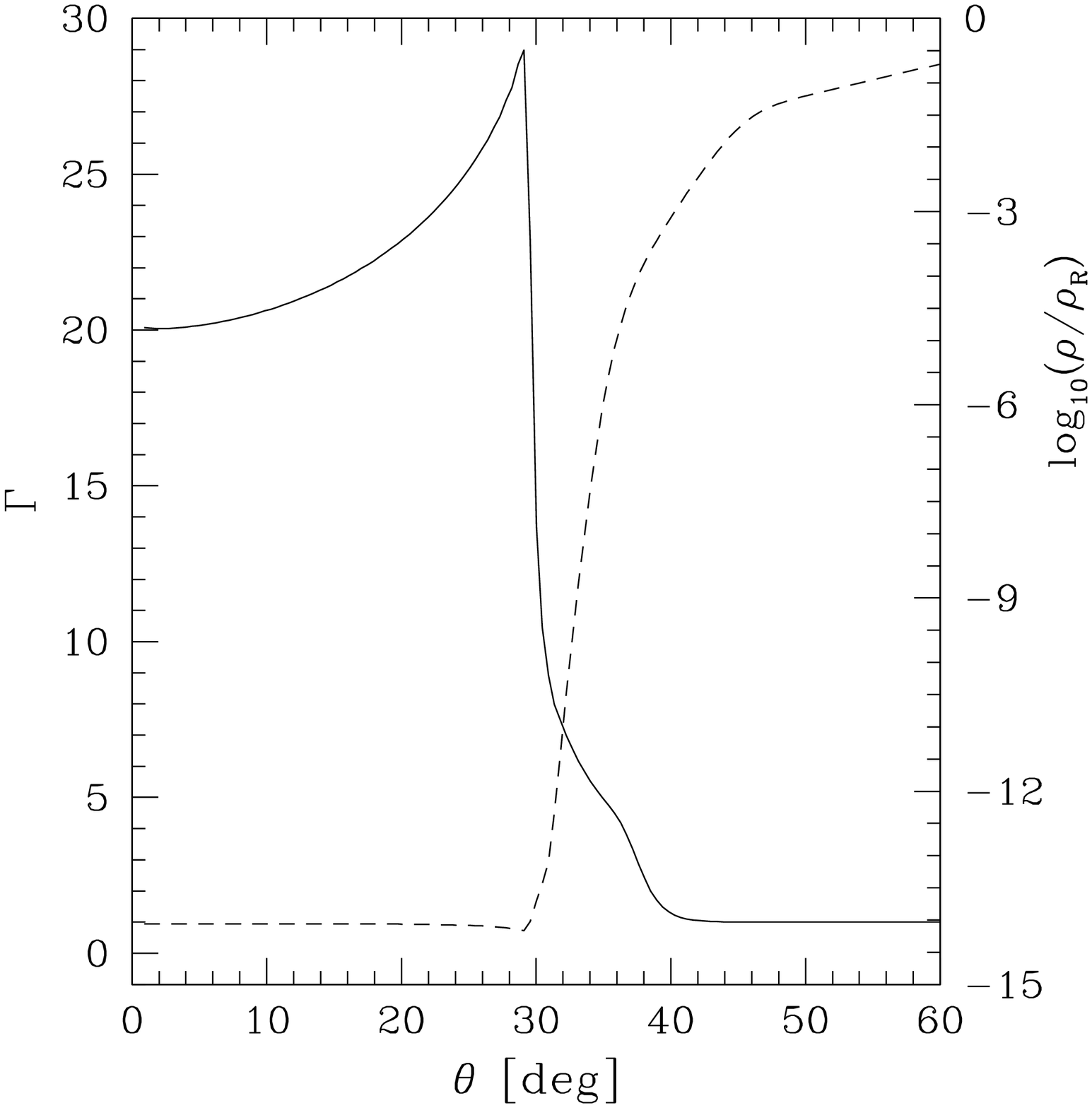}}
\figcaption[]{\label{fig:GRB}Example of the growth of the Lorentz from a
    multidimensional simulations of ultrarelativistic jets generated in
    post-neutron star mergers (data from model A09 of Aloy et
    al. 2005). This is to be compared with the similar results shown in
    the right panel of Fig.~\ref{fig:scs_rcs}.}
\vskip 0.125truecm } 

All of this is summarized in Fig.~\ref{fig:pl_veln}, where the maximum
values of the Lorentz factors are shown as function of the initial value
of the normal velocity in the left state $v^n_{_{\rm L}}$ (Note that the
value of $\Gamma_{_{\rm L}}$ is held fixed so that the increase of
$v^n_{_{\rm L}}$ corresponds to a decrease of $v^t_{_{\rm
L}}$.). Different curves refer to different values of the pressure in the
left state $p_{_{\rm L}}$, with {solid} lines indicating solutions of
the Riemann problem of the type ${\cal _{\leftarrow}\!R C
S_{\!\rightarrow}}$, {dashed} lines to solutions of the type ${\cal
_{\leftarrow}\!S C S_{\!\rightarrow}}$, and filled circles marking the
critical normal velocity {$(v^n_{_{\rm L}})_c$} at which one
wave-pattern transforms into the other.

\section{Astrophysical {Implications}}
\label{aa}

{A} first important astrophysical implication of the relativistic effects
discussed here is that of requiring {\it less extreme} {energy contents}
at the launching sites of relativistic jets and GRBs. This hydrodynamic
mechanism, in fact, can boost a fluid to {ultrarelativistic regimes very
efficiently and thus the Lorentz factors either expected or observed in
astrophysical sources can be explained} with initial ``left-states'' that
have considerably smaller pressures and are thus easier to produce in
practice (in case the jet acceleration is mostly driven by pressure
gradients).

To be more specific, for an ultrarelativistic jet expected in a short
GRB, the conditions close to the generation site ({\em i.e.,} above
the poles of a rapidly rotating black hole) could be associated to a
left state with $p_{_{\rm L}} \sim {(10^{-2}-10)\rho_{_{\rm ext}}}$,
$v^n_{_{\rm L}} \approx 0$, $v^t_{_{\rm L}} \approx {1}$ and
$\rho_{_{\rm L}}/\rho_{_{\rm R}} \ll 1$ [For a short GRB progenitor
$\rho_{\rm ext} \sim 1-10^3\ {\rm g/cm^{3}}$ (Janka et al.  1999)].
In this case, a Lorentz factor as small as $\Gamma_{_{\rm L}} \simeq
10$ is sufficient to produce, through the acceleration across the
rarefaction wave, the inferred Lorentz factors of $\Gamma_{\rm obs}
\gtrsim 100$.  Furthermore, because the pressure in the jet decreases
more rapidly than the pressure in the external medium ({\it i.e.,}
$p_{_{\rm L}} \propto z^{-4}$, $p_{_{\rm R}} \propto z^{-3}$, where
$z$ is the distance from the black hole) the boosting effect will be
confined to a region within which $p_{_{\rm L}}/p_{_{\rm R}} {\gtrsim}
1$. For a black hole with a mass of $\sim 3 M_{\odot}$ this may happen
for $z \lesssim 300\,$km.  {We also note that if the evolution of
  these boundary regions is maintained until the afterglow phase, then
  some differences in the emission properties are to be expected with
  respect to the standard model of the ``universal'' jet-profile ({\it
    i.e.,} a top-hat profile with sharp edges) or of the
  ``structured-jet'' model ({\it i.e.,} where a progressive decay of
  the energy and of the Lorentz factor is expected from the jet core
  to the external medium). The impact of these structural changes
  within the jet on the light curve, whose calculation goes beyond the
  scope of this letter, will help to collect evidence on the
  occurrence of this relativistic acceleration.}

\vbox{
\vskip 0.125truecm
\centerline{\epsfxsize=8.truecm\epsfbox{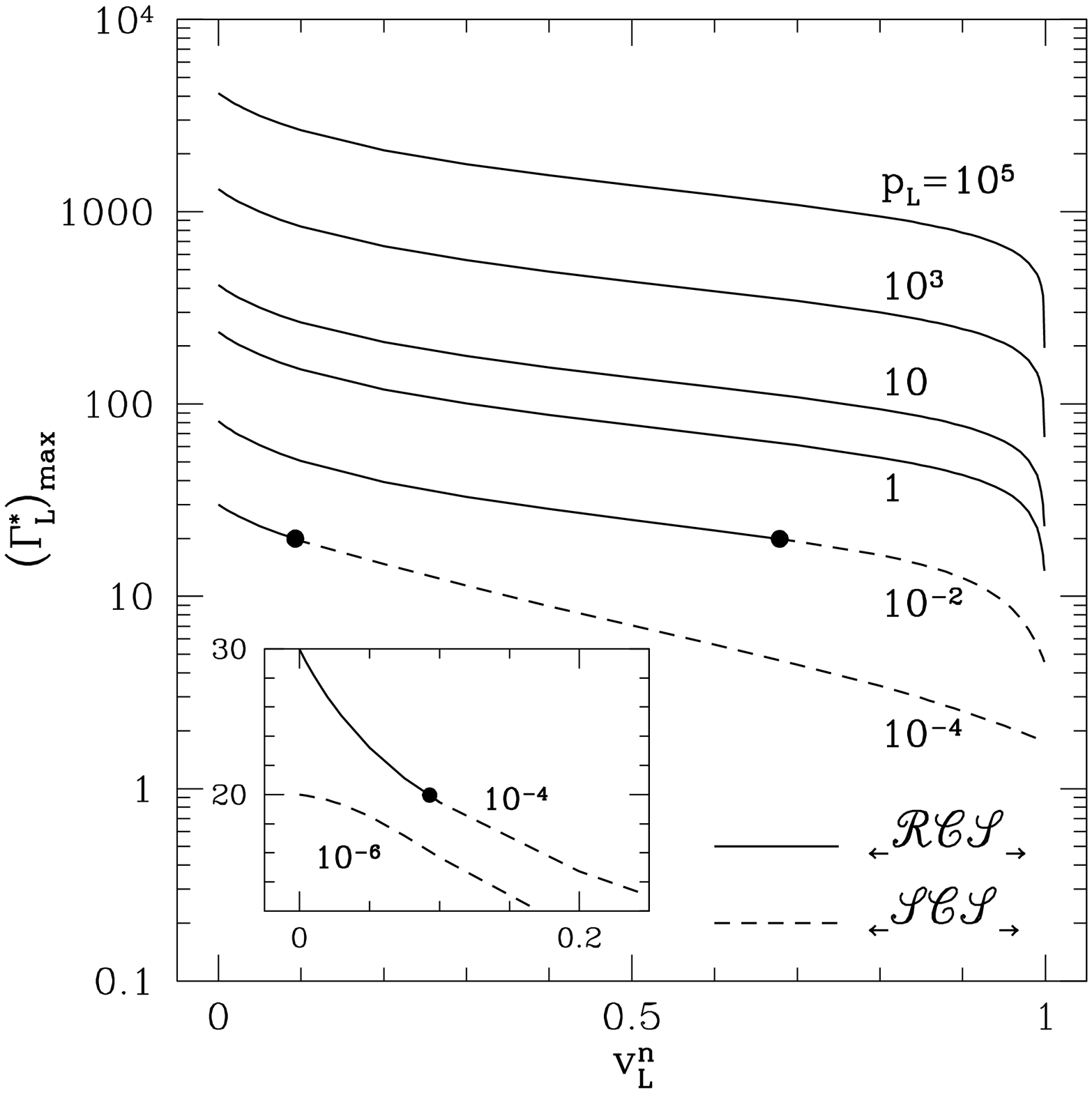}}

\figcaption[]{\label{fig:pl_veln}%
        Maximum Lorentz produced in the Riemann problem $(\Gamma^*_{_{\rm
        L}})_{\rm max}$ as function of the normal velocity in the left
        state $v^n_{_{\rm L}}$. The right state is held fixed and is
        given by $(p_{_{\rm R}}, \rho_{_{\rm R}}, v^n_{_{\rm R}},
        \Gamma_{_{\rm R}}) = (10^{-6},10^{-2},0,1)$. {The left state has
        fixed values in the rest-mass density and Lorentz factor
        $(\rho_{_{\rm L}}, \Gamma_{_{\rm L}}) = (10^{-4},20)$, while the
        pressure $p_{_{\rm L}}$ is varied for each curve as indicated by
        the different labels}. {Continuous lines refer to a ${\cal
        _{\leftarrow}\!R C S_{\!\rightarrow}}$ pattern, while dashed
        lines to a ${\cal _{\leftarrow}\!S C S_{\!\rightarrow}}$;} filled
        circles mark the critical normal velocity $(v^n_{_{\rm
        L}})_c$.}
\vskip 0.125truecm
}

In extragalactic relativistic jets, on the other hand, the Lorentz
factors are more moderate, with $\Gamma_{\rm obs} \lesssim
20$~\citep{VC94} and the physical conditions could then be associated
to a left state with $p_{_{\rm L}} \sim {(10^{-6}-10^{-4})\rho_{_{\rm
      ext}} }$, $v^n_{_{\rm L}} \approx 0$, $v^t_{_{\rm L}} \approx
{1}$ and $\rho_{_{\rm L}}/\rho_{_{\rm R}} \sim {10^{-2}-10^{-5}}$ [For
an extragalactic jet $\rho_{\rm ext} \sim 10^{-27}-10^{-24}\ {\rm
  g/cm^{3}}$ (Ferrari 1998).]. In this case, a Lorentz factor
$\Gamma_{_{\rm L}} \simeq 1.5$ is sufficient to produce the observed
boosts. {In addition, if} {a} {continuous channel of plasma is present
  together with to the detected superluminal ejections}, the physical
conditions in galactic microquasar jets might be similar to the ones
found in extragalactic jets, but with an even smaller inferred value
of the Lorentz factor ($\Gamma_{\rm obs} \lesssim 5$; \citealp{MR99}).
In this case, values as small as $\Gamma_{_{\rm L}} \simeq 1.2$ would
be sufficient to yield the observed apparent speeds.

{A second important astrophysical implication of our results is of
  pointing out that, under suitable but realistic conditions,
  ultrarelativistic regimes can be reached {\it very close} to the jet
  launching site. Furthermore, while the boost originates and is
  larger in a small region at the interface between the jet and the
  external material, it is not restricted to a thin boundary layer but
  can rapidly involve the whole jet. This is because the left-going
  rarefaction wave can sweep up the jet material at relativistic
  speeds and thus rapidly accelerate a considerable mass fraction in
  the jet. As an example, on the basis of the values deduced from the
  Riemann problem considered in the right panel of
  Fig.~\ref{fig:scs_rcs}, more than 10\% of the total (baryonic) mass
  is accelerated to $\Gamma > \Gamma_{_{\rm L}}$ in less that 1/5 of
  the} {lateral jet-crossing time.} In contrast, in other thermal
  acceleration mechanisms that also exploit the existence of $h \gg 1$
  ({\it e.g.,} the fireball model for GRBs) some time is needed to
  speed-up the flow, with ultrarelativistic regimes being reached far
  from the launching site.

        It is important to underline that the boosting mechanism
proposed here should not be considered an alternative to the fireball
model. Rather, it complements it by providing a more detailed description
of the relativistic flows at the launching sites. Indeed, we expect both
processes of hydrodynamic acceleration to operate simultaneously, with a
longitudinal acceleration being produced by the expansion of a hot gas,
and a the lateral one being produced by the relativistic effects
discussed here.

{Interestingly, this boosting mechanism} can also enhance particle
acceleration when a turbulent velocity shear layer develops along a
parsec-scale jet. These layers, in fact, whose existence has been pointed
out both observationally ({\it e.g.,}~\citealp{OHC89, SBB98}) and through
numerical simulations~\citep{Alo99b,Alo00}, represent natural sites for
particle acceleration, providing high-energy cosmic rays and influencing
the dynamics of relativistic jets in extragalactic radio sources by
forming cosmic-ray cocoons~\citep{Ostrowski00}. The efficiency of the
acceleration process in these turbulent shear layers depends on the
particle mean-free-path and on the velocity structure~\citep{SO02}. Our
results suggest that particle production can be further amplified if the
simple flow structure usually assumed in these calculations and similar
to the one produced in a ${\cal _{\leftarrow}\!S C S_{\!\rightarrow}}$ is
instead replaced by a more realistic one, such as the one occurring in a
${\cal _{\leftarrow}\!R C S_{\!\rightarrow}}$.

As a final remark, we point out that the large hydrodynamic boosts
{reported} here may be detected with laboratory experiments
involving heavy-ion collisions. In these experiments, heavy ions are
accelerated to ultrarelativistic velocities and collided. If the two
beams are chosen to have different specific enthalpies and to collide
with a nonzero impact parameter, an acceleration of the interacting
layer could be produced as a result of the relativistic boosting
discussed here.

\begin{acknowledgments}
  MAA is a Ram\'on y Cajal Fellow of the Spanish MEC.  Support to this
  research comes also through the Spanish {MEC} (AYA2004-08067-C03-C01)
  and the SFB-TR7 ``Gravitationswellenastronomie'' of the DFG.
\end{acknowledgments}



\end{document}